\documentstyle[12pt]{article}
\def\beq{\begin{equation}}
\def\brr{\begin{array}}
\def\err{\end{array}}
\def\eeq{\end{equation}}
\def\bea{\begin{eqnarray}}
\def\eea{\end{eqnarray}}

\def\tr{\mbox{Tr}\, }
\def\ni{\noindent}

\def\nn{\nonumber}
\def\ms{\medskip}

\begin{document}

\hfill UB-ECM-PF 92/34

\hfill December 1992

\vspace*{3mm}

\begin{center}

{\LARGE \bf
A finite-temperature periodic structure  in (super)string theory }

\vspace{4mm}


{\sc A.A. Bytsenko} \\ {\it Department of Theoretical Physics, State
Technical
University, \\ St Petersburg 195251, Russia} \\
{\sc E. Elizalde}\footnote{E-mail address: eli @ ebubecm1.bitnet}
\\
{\it Department E.C.M., Faculty of Physics, University of
Barcelona, \\
Diagonal 647, 08028 Barcelona, Spain} \\
{\sc S.D. Odintsov}\footnote{On sabbatical leave from
Tomsk Pedagogical Institute, 634041 Tomsk, Russia.} \\
{\it Department of Physics, Faculty of Science, Hiroshima
University, \\
Higashi-Hiroshima 724, Japan} \\
 {\sc S. Zerbini} \\
{\it Department of Physics, University of Trento, 38050 Povo,
Italy}    \\
{\it I.N.F.N., Gruppo Collegato di Trento}
\ms


\vspace{5mm}

{\bf Abstract}

\end{center}

Using a Laurent series representation for the (super)string one-loop
free energy,
 an explicit form for the analytic continuation of
the Laurent series beyond the critical (Hagedorn) temperature is
obtained. As an additional result, a periodic structure is found in
(super)string
thermodynamics. A brief physical discussion about the origin and
meaning of such  structure is carried out.



\newpage

\section{Introduction}

There is a number of motivations for the study  of the
 behaviour of extended objects ---here we will mainly concentrate on
strings--- at
non-zero temperature $T=\beta^{-1}$ (for a short list of references see
[1-6]).
Perhaps one
of the main reasons for these investigations is connected with the
 thermodynamics  of the
 early universe (see [7] and references therein) as well as with
the attempts  to use exented objects  for the description of the
high temperature limit of
the confining phase of large-N SU(N) Yang-Mills theory [8,9].

It is well known since the early days of dual string models, that an
essential ingredient of string theory at non-zero temperature is
the so-called Hagedorn temperature.
The appearance of the Hagedorn temperature is a consequence of the
fact that the asymptotic form of the state level density has an
exponential dependence of the mass. A naive argument leads to the
conclusion that above such a temperature the free energy
diverges.
According to the popular viewpoint, the Hagedorn temperature
is the critical temperature for a phase transition to some new
phase (probably associated with topological strings [10]).

There are different representations (in particular, those connected
with different ensembles) for the string free energy. One of these
representations [6] ---which is very useful for formal
manipulations--- gives a modular-invariant expression for the free
energy. However, this and all the other well known representations
 [5] are {\it integral} ones, in which the Hagedorn
temperature appears as the  convergence condition in the
ultraviolet limit of a certain integral.
In order to
discuss the high- or the low-temperature limits in such
representations one must expand the integral in terms of a
corresponding series. Thus, a specific series expansion
appears in string theory at non-zero temperature.

In the present note, making use of the so-called Laurent series
representation for the one-loop open superstring free energy
introduced in refs. [11,12] and of the analytic
continuation of such series, we discuss the possible appearance of a periodic
thermodynamic structure valid for any value of $\beta$. Some consideration
about the open bosonic string are also reported.

\section{One-loop (super)string free energy at finite temperature}
 It is  well-known that the one-loop
free energy for the bosonic ($b$) or fermionic ($f$) degree of
freedom in $d$-dimensional space is given by
\beq
F_{b,f} = \pm \frac{1}{\beta} \int \frac{d^{d-1} k}{(2\pi )^{d-1}}
\log \left(1\mp  e^{-\beta \omega_k} \right)
\eeq
where $\omega_k=\sqrt{k^2+m^2}$ and
$m$ is the mass for the corresponding degree of freedom.
Let us recall the Mellin-Barnes representation for the one-loop free
energy  discussed in field theory in ref.
[13] (see also [14,15]).
Integrating eq. (1) by parts, we obtain
\beq
F_{b,f} =- \frac{(4\pi)^{(1-d)/2}}{(d-1)\Gamma ((d-1)/2)} \int dk^2
\,  \frac{k^{d-1}}{\omega_k \left( e^{\beta \omega_k} \mp 1 \right)}.
\eeq
For the factor in the integrand, $\left( e^{\beta \omega_k} \mp 1
\right)^{-1}$, we shall use the Mellin transform in the following
form [13]
\beq
 \frac{1}{ e^{ax} \mp 1 } = \frac{1}{2\pi i} \int_{c-
i\infty}^{c+i\infty} ds \,  \zeta^{(\mp)} (s) \Gamma (s)
(ax)^{-s},     \ \  \ \ \ \ ax >0,
\eeq
where Re $s=c, \ c>1$, for bosons and $c>0$ for fermions, $
\zeta^{(-)} (s)=\zeta (s)$ is the Riemann-Hurwitz zeta function and
$ \zeta^{(+)} (s)=\left( 1-2^{1-s} \right) \zeta (s)$ $=
\sum_{n=1}^{\infty} (-1)^{n-1} n^{-s}$, Re $s>0$.

Substituting (3) into (2), we get
 \beq
F_{b,f} =- \frac{(4\pi)^{(1-d)/2}}{(d-1)\Gamma ((d-1)/2)} \int dk^2
\, k^{d-1} \frac{1}{2\pi i} \int_{c-i\infty}^{c+i\infty} ds \,
\zeta^{(\mp)} (s) \Gamma (s) \beta^{-s} \omega_k^{-1-s}.
\eeq
Performing the integration over $k$ with the help of the Euler beta function
$B(x,y)=\Gamma (x)\Gamma (y)/\Gamma (x+y)$ (notice that, owing to
absolute convergence, the order of integration over $k$ and $s$ can
be interchanged), we obtain
\beq
 \int dk^2 \, k^{d-1}  \omega_k^{-1-s} = (m^2)^{(d-s)/2} B \left(
\frac{d+1}{2}, \frac{s-d}{2} \right) = (m^2)^{(d-s)/2} \frac{
\Gamma \left( \frac{d+1}{2} \right) \Gamma \left( \frac{s-d}{2}
\right)}{ \Gamma \left( \frac{s+1}{2} \right)}, \ \ \mbox{Re} \, s
>d.
\eeq
Finally, one gets [13]

\beq
F_{b,f} =- 2^{-d} \pi^{(1-d)/2} \frac{1}{2\pi i} \int_{c-
i\infty}^{c+i\infty} ds \,  \Gamma (\frac{s}{2})
\zeta^{(\mp)} (s)  {\frac{\beta}{2}}^{-s}  \Gamma \left(
\frac{s-d}{2} \right) (m^2)^{(d-s)/2},  \ \ \
c >d.
\eeq
 For (super)string theory one can write the representation (6) in the form
\bea
F_{\mbox{bosonic string}} &=&- 2^{-d-1} \pi^{-d/2} \frac{1}{2\pi
i} \int_{c-i\infty}^{c+i\infty} ds \,\zeta (s)  (\frac{\beta}{2})^{-s}
\Gamma (\frac{s}{2})
\Gamma \left( \frac{s-d}{2} \right)  \nn \\
 &&  \times  \tr (M^2)^{(d-s)/2} \ , \ \ \ \ d=26\ \ \ ,
\eea
\bea
F_{\mbox{superstring}} &=&- 2^{-d} \pi^{-d/2} \frac{1}{2\pi i}
\int_{c-i\infty}^{c+i\infty} ds \, \zeta (s)  (1-2^{-s})
(\frac{\beta}{2})^{-s}
 \Gamma (\frac{s}{2}) \Gamma
\left( \frac{s-d}{2} \right)
 \nn \\
&& \times \mbox{STr} \, (M^2)^{(d-s)/2}, \ \ \ \ \
d=10
\ \ \ .
\eea
Here $M^2$ is the mass operator, the symbol $\mbox{STr}$ means
the trace over fermion and boson fields. For closed strings the constraint
should be introduced via the usual identity [2].

For the bosonic strings the mass operator contains both infrared (due
to the presence of the tachyon in the spectrum) and ultraviolet
divergences, while for superstrings it contains only ultraviolet
divergences. Hence, the consideration of superstrings is much
simpler from a technical point of view.

In the following we shall consider open
superstrings.  For the open superstring (without gauge
group) the spectrum is given by (see for example [16])
\beq
M^2=2\sum_{i=1}^{d-2}\sum_{n=1}^{\infty}  n \left( N_{ni}^b +
N_{ni}^f \right), \ \ \ \ d=10.
\eeq
The quantity $\mbox{STr} (M^2)^{(d-s)/2}$ which appears in equation (8),
requires a
regularization because a
naive definition of it leads to a formal divergent expression, namely
at $d=10$ we can write
\beq
 \mbox{STr} \, (M^2)^{5- \frac{s}{2}} =  \frac{1}{\Gamma
\left( \frac{s}{2}-5 \right)} \int_0^{\infty} dt \, t^{
\frac{s}{2} -6} \mbox{STr} e^{-tM^2} .
\eeq
As a result we need
the heat-kernel expansion for $ \mbox{STr} \, e^{-tM^2}$.

It is known that
\beq
 \mbox{STr} \, e^{-tM^2} = 8 \prod_{n=1}^{\infty} \left( \frac{1-e^{-
2tn}}{1+e^{-2tn}} \right)^{-8} = 8 \left[ \theta_4 \left( 0 |
e^{-2t} \right)\right]^{-8},
\eeq
the presence of the factor $8$ in eq. (11) is due to the degeneracy of
the ground state.
Recall that the Jacobi's elliptic theta function
$\theta_4 \left( 0 | e^{-t} \right) $ when $ t \rightarrow 0$ has the
asymptotic form
\bea
\theta_4 \left( 0 | e^{-t} \right) &=&
\sqrt{\frac{\pi}{t}} \,
\theta_2 \left( 0 | e^{-\pi^2/t} \right) =
\sqrt{\frac{\pi}{t}}
\sum_{n=-\infty}^{+\infty} \exp \left[ -\frac{\pi^2}{t} \left( n-
\frac{1}{2} \right)^2 \right] \nn \\
&=& 2
\sqrt{\frac{\pi}{t}}
 \, e^{-\pi^2/(4t)} \left( 1+ e^{-2\pi^2/t}+ e^{-6\pi^2/t} +
\cdots \right), \ \
\eea
hence
\beq
\left. \left[ \theta_4 \left( 0 | e^{-t} \right)\right]^{-
8} \right|_{t\rightarrow 0} = \frac{t^4}{2^8\pi^4} \, e^{2\pi^2/t}
-\frac{t^4}{2^5\pi^4} + {\cal O} \left( e^{-2\pi^2/t} \right).
\eeq
So we can define the regularized supertrace in the following
way
\bea
 \mbox{STr} \, (M^2)^{5- \frac{s}{2}}& =&  \frac{2^{8-s/2}}{\Gamma
\left( \frac{s}{2}-5 \right)} \left\{ \int_0^{\infty} dt \, t^{
\frac{s}{2} -6} \left[ \left[ \theta_4 \left( 0 | e^{-t}
\right)\right]^{-8} - \frac{t^4}{2^8\pi^4} \, \left( e^{2\pi^2/t}
-
8 \right) \right] \right. \nn \\
& +&\left.   \frac{1}{2^8\pi^4} \, \int_0^{\infty} dt \, t^{
\frac{s}{2} -2} \left( e^{2\pi^2/t} -8 \right)
\right\}.
\eea
Since the regularization of the integral $ \int_0^{\infty} dx \,
x^{\lambda}$ as the analytical function of $\lambda$ gives $
\int_0^{\infty} dx \, x^{\lambda} =0$, it turns out that the last
integral in eq. (14) is equal to zero. Moreover, we have
\beq
 \int_0^{\infty} dt \, t^{ \frac{s}{2} -2} e^{2\pi^2/t}
=   \int_0^{\infty} dt \, t^{\left(- \frac{s}{2}
+1\right)-1}  e^{2\pi^2t} =(- 2\pi^2)^{ \frac{s}{2} -1} \Gamma
\left(1-\frac{s}{2} \right),
\eeq
and therefore
\beq
 \mbox{STr} \, (M^2)^{5- \frac{s}{2}} = \frac{2^{-1}\pi^{-
6}}{\Gamma \left(
\frac{s}{2}-5 \right)} \left[ \pi^s \Gamma \left(1-\frac{s}{2}
\right) \mbox{Re} \,  (-1)^{\frac{s}{2} -1}  +2 \pi^{3/2} G(s,\mu)
\right],
\eeq
where
\beq
G(s, \mu)= 2^{1-s/2} \pi^{(s-1)/2}  \int_0^{\mu} dt
\, t^{ \frac{s}{2} -6}
\left\{ \left[ \frac{1}{2} \theta_4 \left( 0 | e^{-\pi t}
\right)\right]^{- 8} -  t^4\left( e^{2\pi/t}-8 \right) \right\}.
\eeq
In eq. (17) the infrared cutoff parameter $\mu$  has been introduced.
Such a regularization is necessary  for $s\geq 4$. On the next
stage of our calculations this regularization will be removed ($\mu
\rightarrow \infty$).

For the one-loop free energy we get
\beq
F_{\mbox{superstring}} = - (2 \pi)^{-11} \frac{1}{2\pi i}
\int_{c-i\infty}^{c+i\infty} ds \, \left[ \varphi (s) + \psi (s)
\right],
\eeq
where
\bea
\varphi (s) &=& (1-2^{-s} )  [\mbox{Re} \, (-1)^{\frac{s}{2} -1}]
 \frac{ \pi}{\sin \frac{\pi s}{2}} \zeta (s)
\left( \frac{\beta}{2\pi}
\right)^{-s}, \\
\psi (s) &=&  (1-2^{-s} ) (\pi)^{3/2-s} G(s,\mu) \,\Gamma(s/2)
 \zeta (s)
(\frac{\beta}{2 \pi})^{-s}.
\eea

The meromorphic function $\varphi (s)$ has first order poles at
$s=1$ and $s=2k$, $k=0,1,2, \ldots$.
The pole of $\varphi$ for $s=1$  is also of first order but its
residue
is imaginary. The meromorphic function $\psi
(s)$ has first order poles at $s=1$.  One can see that the
regularization cutoff parameter is removed
automatically.

As a result, we obtain
\beq
F_{\mbox{superstring}}  = 2(2 \pi)^{-11}  \left[
\sum_{k=1}^{\infty}(1-2^{-2k})
\zeta(2k) x^{2k} -\frac{\pi x}{4} G(1, \infty)
\right]+ F_R(x),
\eeq
where $ x=\beta_c/\beta$ , $\beta_c=2\pi$ and $F_R(x)$ is the contribution
 coming from the
contour integral along the arc of radius $R$. If $|x|<1$, then the value
of the contour integral on the right half-plane
is vanishing when $R \rightarrow \infty$.
Therefore the series converges
when $\beta > \beta_c =2\pi$, $\beta_c$ being the Hagedorn temperature
[1-4].
 The sum of the series can be
explicitly evaluated and the result is
\beq
\sum_{k=1}^{\infty}(1-2^{-2k})
\zeta(2k) x^{2k}=\frac{\pi x}{4}\tan (\frac{\pi x}{2}),\ \  |x|<1  .
\eeq
As a consequence, the free energy is given by

\beq
F_{\mbox{superstring}}  = \frac{\pi x}{2(2 \pi)^{11}}  \left
[ \tan (\frac{\pi x}{2}) -G(1, \infty)
\right]  .
\eeq

Now, observing that
\beq
\sum_{k=1}^{\infty}
\zeta(2k) y^{2k}=\frac{1}{2}-\frac{\pi y}{2}\cot (\pi y),\ \  |y|<1
 ,
\eeq
it is easy to
show that  the free energy for the open bosonic string has the form
\beq
F_{\mbox{bosonic string}}  = \frac{1}{2^{23} \pi^{16}} \left
[-y\cot (\pi y) -\frac{2y}{ \beta_c}D(1,\mu)+
\pi^{-1} D(0,\mu) \right].
\eeq
Here $y=\beta_c/\beta$ and the related Hagedorn temperature is
$\beta_c= \sqrt 8 \pi $,
\beq
D(s,\mu)=2 \pi \int_0^{\mu} dt t^{s/2-14}\left [ \eta(it)^{-24}-
t^{12} e^{2\pi /t} \right] ,
\eeq
and $ \eta(\tau)= e^{i\pi \tau /12} \prod_{n=1}^{\infty} (1-e^{2\pi i n
\tau})$ is Dedekind's eta function.
 However, in this case, the infared cutoff
parameter $\mu$ cannot be
removed (there is a tachion in the spectrum).

The Laurent series  have been obtained in each case for
for $|x|<1$ and $|y|<1$, respectively, namely for $\beta > \beta_c$.
However the right hand sides of the above formulas (22) and (24)  may
also be understood as analytic continuations of these  series to
arguments $|x|>1$  and $|y|>1$, respectively (i.e. to $\beta <
\beta_c$).
As a consequence, we realize that we have obtained a periodic structure
for the one-loop free energy of the corresponding (super)strings.

\section{Conclusions}
We finish the paper with some remarks. The results we have obtained
here are based on the Mellin-Barnes representation for the one-loop
free energy of the critical (super)strings. Such a novel
representation for the lowest order in string perturbation theory has
allowed us to obtain  explicit thermodynamic expressions in term of a
Laurent series. The critical temperature arises in this formalism as
the convergence condition (namely, the radius of convergence) of these
series. Furthermore, an explicit analytic continuation of the free
energy to temperatures beyond the critical Hagedorn temperature (i.e.,
to $ \beta
\leq \beta_c$) has been constructed. As a result, there is now the
possibility to have the free energy which corresponds to new string
phases.

Actually, it is somewhat surprising to see that there exists such a
finite temperature
periodic stucture in the behaviour of (super)string thermodynamical
quantities. However, we should remember  that a  similar asymptotic
behaviour was obtained in ref. [1].
So there is a possibility for a new interpretation of the critical
temperature and of the transitions associated with this thermodinamical
structure.

In the case of the closed bosonic string ---which has not
been considered here--- these transitions may well be connected with the
dual
$\beta$-symmetry which is present in the function $ \tr \exp (-tM^2)$
and hidden in the final form  of the free energy.

The typical widths of the periodic sectors depend on the Regge slope
parameter $\alpha $. The widths of the sectors grow together with the
parameter $\alpha$, and in the limit $\alpha \rightarrow \infty$
(string tension goes to zero), the thermodynamic system can be
associated with an ideal gas of  free quantum fields present in the
normal modes of the string [2,17]. Of course the precise interpretation of
``domain"-like structure in quantum (super)string theory obtained in
this paper is somehow delicate and many questions remain unanswered.

Finally it should be noticed that higher orders of string perturbation
theory do not modify the critical temperature, at least for the
bosonic string [18]. In consequence, there exists the possibility
that the periodic structure found above might also be present when
 dealing with a Riemann surface of arbitrary genus.

\vspace{2mm}

\ni{\large \bf Acknowledgments}

We would like to thank G. Cognola, K. Kirsten and L. Vanzo for
discussions. A.A.B. is grateful to the Faculty of Science of the University of
Trento for hospitality. E.E. thanks DGICYT (Spain) for finantial support
through research project PB90-0022.
 S.D.O. wishes to thank the Particle Group at
Hiroshima University
for kind hospitality.


\end{document}